\documentclass[10pt]{amsart}

\usepackage{tikz}
\usepackage{amssymb,amsmath,latexsym,graphicx}
\usepackage{charter,eucal}
\usepackage{amssymb}
\usepackage{amscd}

\newtheorem{theorem}{Theorem }[section]
\newtheorem{lemma}[theorem]{Lemma}
\newtheorem{remark}[theorem]{Remark}
\newtheorem{corollary}[theorem]{Corollary}
\newtheorem{proposition}[theorem]{Proposition}
\newtheorem{question}[theorem]{\textsc{Question}}
\newtheorem{conjecture}[theorem]{\textsc{Conjecture}}
\newtheorem{definition}[theorem]{\textsc{Definition}}

\def\I{\mathrel{\mathrm{I}}}

\def\1{\mathrel{\mathbf{1}}}

\newcommand{\mS}{\mathcal{S}}
\newcommand{\mT}{\mathcal{T}}

\newcommand{\mG}{\mathcal{G}} 
\newcommand{\mR}{\mathcal{R}}

\newcommand{\mQ}{\mathcal{Q}} 

\newcommand{\bI}{\mathbf{I}}

\newcommand{\mO}{\mathcal{O}}

\newcommand{\eop}{\hspace*{\fill}$\blacksquare$}

\newcommand{\btt}{\begin{ttheorem}}
\newcommand{\ett}{\end{ttheorem}}

\newcommand{\bt}{\begin{theorem}}
\newcommand{\et}{\end{theorem}}
\newcommand{\bcc}{\begin{conjecture}}
\newcommand{\ecc}{\end{conjecture}}
\newcommand{\bc}{\begin{corollary}}
\newcommand{\bl}{\begin{lemma}}
\newcommand{\ec}{\end{corollary}}
\newcommand{\el}{\end{lemma}}
\newcommand{\bq}{\begin{question}}
\newcommand{\eq}{\end{question}}
\newcommand{\bp}{\begin{proposition}}
\newcommand{\ep}{\end{proposition}}
\newcommand{\br}{\begin{remark}}
\newcommand{\er}{\end{remark}}
\newcommand{\bd}{\begin{definition}}
\newcommand{\ed}{\end{definition}}

\newcommand{\mW}{\ensuremath{\mathcal{W}}}

\newcommand{\PG}{\ensuremath{\mathbf{PG}}}

\newcommand{\mE}{\ensuremath{\mathcal{E}}}
\newcommand{\mL}{\ensuremath{\mathcal{L}}}

\newcommand{\mC}{\ensuremath{\mathcal{C}}}
\newcommand{\mX}{\ensuremath{\mathcal{X}}}

\newcommand{\hP}{\ensuremath{\mathbf{P}}}
\newcommand{\mU}{\ensuremath{\mathcal{U}}}

\newcommand{\mB}{\mathcal{B}}
\newcommand{\mP}{\mathcal{P}}

\author{Koen  Thas}

\address{Department of Mathematics,
Ghent University,
Krijgslaan 281, S25, B-9000 Ghent, Belgium}

\email{koen.thas@gmail.com}

\title[Unextendible mutually unbiased bases]{Unextendible mutually unbiased bases\\ (after Mandayam, Bandyopadhyay, Grassl and Wootters)}

\date{}

\begin{document}

\maketitle

\begin{abstract}
We consider  questions posed in a recent paper of Mandayam, Bandyopadhyay, Grassl and Wootters \cite{Woot}
on the nature of ``unextendible mutually unbiased bases.'' 
We describe a conceptual framework to study these questions,
using a connection proved by the author in \cite{Appl}
between the set of nonidentity generalized Pauli operators on the Hilbert space
of $N$ $d$-level quantum systems, $d$ a prime, and the geometry of non-degenerate alternating bilinear forms of rank $N$ over finite fields $\mathbb{F}_d$.
We then supply alternative and short proofs  of results obtained in \cite{Woot}, as well as new general bounds for the problems considered in {\em loc. cit}.\\ 
In this setting, we also solve Conjecture 1 of \cite{Woot},  and speculate on variations of this conjecture.
\end{abstract}

{\bf PACs numbers}: 02.10.Hh, 02.40.Dr, 03.67.-a, 03.65.Ta, 03.65.Ud\\

\setcounter{tocdepth}{1}
\tableofcontents

\bigskip
\section{Introduction}
Finite-dimensional quantum systems | that is, ``multiple qudits'' |
exhibit many interesting properties like quantum entanglement and
quantum non-locality and play, therefore, a crucial role in
numerous physical applications like Quantum Cryptography, Quantum
Coding, Quantum Cloning/Teleportation and/or Quantum Computing, to
mention just a few. As these systems live in
finite-dimensional Hilbert spaces, further insights into their
behavior require, obviously, a proper understanding of the
structure of the associated Hilbert spaces. Within the past few
years, a lot of activity in this direction has been devoted to the
study of so-called mutually unbiased bases (``MUBs'').\\


Recall that two orthonormal bases $\mB$ and $\mB'$ of the Hilbert space $\mathbb{C}^\ell$ ($\ell \in \mathbb{N}^{\times}$) are {\em mutually unbiased}
if and only if

\[ \vert \langle \phi \vert \psi \rangle \vert^2 = 1/\ell            \]
\noindent
for all $\vert\phi\rangle \in \mB$ and $\vert\psi\rangle \in \mB'$.
It is a fundamental conjecture, with many applications, that the theoretical upper bound 
$\ell + 1$ of a set of mutually unbiased bases  can only be reached when $\ell$ is a {\em prime power}.\\

It has been suspected for a long time that there are deep connections between Quantum (Information) Theory
and Finite Geometry | see, for instance, Wootters \cite{woo,woo-fg}. (See also \cite{HS1}--\cite{HS2} and \cite{san}--\cite{SaPl}, and references therein.)

As a specific example, proving a conjecture of Saniga and Planat \cite{SaPl},  the author showed in
\cite{Appl}  that the generalized Pauli operators can be identified with the points, and maximum sets of pairwise commuting members of them with the lines (or subspaces of higher dimensions), of a specific finite incidence geometry so that the structure of the operator space can fully be inferred from the properties of the geometry in question. The incidence geometry is the geometry of a non-degenerate alternating bilinear form over a finite field, called {\em symplectic polar space}. Using this connection, it is easy to construct maximal sets of MUBs by just translating known results in the theory of symplectic polar spaces.\\

In a recent paper \cite{Woot},
Mandayam, Bandyopadhyay, Grassl and Wootters 
introduced {\em unextendible mutually unbiased bases} (``UMUBs'') (and several variations and related concepts; details can be found in \S \ref{UMUBdef}) as a natural generalization of maximal sets of mutually unbiased bases.

One of the main results of \cite{Woot} reads as follows.

\bt[Mandayam et al. \cite{Woot}]
\label{Wootmain1}
Given three Pauli classes $\mC_1,\mC_2,\mC_3$ belonging to a complete set $\mS$ of classes in $d = 4$, there exists exactly one more maximal
commuting class $\mC$ of Pauli operators in $\mC_1 \cup \mC_2 \cup \mC_3$.  The class $\mC$ together with the remaining two classes $\mC_4$ and $\mC_5$ of $\mS$ forms an unextendible set of Pauli classes, whose common eigenbases form a weakly UMUB of order $3$.
\et

Using the connection with the polar space, we will give a short proof of this result. Moreover, we generalize this result
for all dimensions $\ell =$ prime$^2$. (In fact, we present a construction of a new class of maximal partial spreads of the symplectic polar space $\mW_3(\ell)$ for any odd prime power $\ell$,  which translates to UMUBs in the case $\ell$ is a prime.)
In dimension $\ell = 8$, a similar result is obtained in \cite{Woot}.

Motivated by Theorem \ref{Wootmain1} and the result in dimension $8$, the following conjecture is then stated in \cite{Woot}.

\bcc[Mandayam et al. \cite{Woot}]
\label{Wootconj}
Given $\ell/2 + 1$ maximal commuting Pauli classes $\mC_1,\mC_2,\ldots,\mC_{\ell/2 + 1}$ belonging to a complete set $\mS$ of classes in $\ell = 2^n =: d^n$, there exists exactly one more maximal
commuting class $\mC$ of Pauli operators in $\cup_{1 \leq i \leq \ell/2 + 1}\mC_i$.  The class $\mC$ together with the remaining classes of $\mS$  forms an unextendible set of Pauli classes of size $\ell/2 + 1$, whose common eigenbases form a weakly UMUB of order $\ell/2 + 1$.
\ecc

We will show that this conjecture is true {\em if and only if $n = 2$ or $n = 8$}. In fact, we will consider the conjecture in any characteristic $d$ (i.e., for any prime $d$), and show that it is true if and only if $d = 2$ and $n = 2$ or $n = 8$.

We then indicate that an alternative version of the conjecture might be true, and describe several new construction techniques to obtain  weakly unextendible sets of MUBs. 

At the end of the paper, we discuss a special kind of weakly unextendible sets of MUBs, called ``Galois MUBs,'' which attain an optimal bound in relation to being unextendible.

\bigskip
{\bf Acknowledgement}.\quad
The author wishes to thank Marcus Grassl and William K. Wootters for various interesting communications on the subject 
of this note.\\

\bigskip
\section{The general Pauli group}

Let $d$ be an odd prime.
Let $\{ \vert s \rangle \vert s = 0,1,\ldots,d - 1\}$ be a computational base of $\mathbb{C}^d$.
Define the $d^2$ (generalized) {\em Pauli operators} of $\mathbb{C}^d$ as

\[  (X_d)^a(Z_d)^b,\ \ \ a,b \in \{0,1,\ldots,d - 1\},         \]
\noindent
where $X_d$ and $Z_d$ are defined by the following actions

\[  X_d \vert s\rangle = \vert s + 1\mod{d}\rangle, \ \ \ Z_d\vert s\rangle = \omega^s \vert s\rangle,        \]
\noindent
where {$\omega =$ exp$(2i\pi/d)$}.

\medskip
The set $\mathbb{P}$ of generalized Pauli operators of the $N$-qudit Hilbert space $\mathbb{C}^{d^N}$ is the set $\mathbb{P}$ of $d^{2N}$ distinct tensor products of the form

\[ \sigma_{i_1} \otimes \sigma_{i_2} \otimes \cdots \otimes \sigma_{i_N},             \]
\noindent
where the $\sigma_{i_k}$ run over the set of (generalized) Pauli matrices of $\mathbb{C}^d$. Denote $\mathbb{P}^{\times} = \mathbb{P} \setminus \{\bI\}$.
These operators generate a group $\hP = \hP_N(d)$ | the {\em general Pauli group} or {\em {discrete} Heisenberg-Weyl group} |
under ordinary matrix multiplication, which has order $d^{2N + 1}$.

For the case of $N$-qubit Hilbert spaces, we refer the reader to \cite{Appl} | it is completely similar.\\

\bigskip
\section{Unextendible sets of MUBs and operator classes}
\label{UMUBdef}

Let $\mU$ be a set of $d^2$ mutually orthogonal unitary operators in $\mathbb{C}^d$ using the Hilbert-Schmidt norm: operators $A$ and $B$ are {\em orthogonal} if $\mathrm{tr}(AB^{\dagger}) = 0$. 
Along with the identity operator $\bI$, $\mU$ constitutes a basis for the $\mathbb{C}$-vector space of $(d \times d)$-complex matrices $\mathbf{M}_{d\times d}(\mathbb{C})$.
A standard construction of MUBs outlined in \cite{Band} relies on finding classes of commuting
operators, with each class containing $d - 1$ mutually orthogonal commuting unitary matrices
different from the identity $\mathbf{I}$.

\subsection{Maximal commuting operator classes} 

A set of subsets $\{\mC_1, \mC_2,\ldots, \mC_{\ell} \vert \mC_j \subset \mU \setminus \{\bI\}\}$ of size 
$\vert \mC_j \vert = d - 1$ constitutes a (partial) partitioning of $\mU \setminus \{\bI\}$ into {\em mutually
disjoint maximal commuting classes} if the subsets $\mC_j$ are such that 
\begin{itemize}
\item[(a)]
the elements of $\mC_j$
commute for all $1 \leq j \leq \ell$ and 
\item[(b)] 
$\mC_j \cap \mC_k = \emptyset$ for all $j \ne k$.
\end{itemize}

In the rest of the paper,  we use the term ``Pauli classes'' to refer to
mutually disjoint maximal commuting classes formed out of the $N$-qudit Pauli group $\mathbf{P}_N(d) \leq \mathbf{U}_{d^N}(\mathbb{C})$.\footnote{In \cite{Woot}, only qubits are considered.}
The correspondence between maximal commuting operator classes and MUBs is stated in
the following lemma, originally proved in \cite{Band}.

\bl[\cite{Band}]
The common eigenbases of $\ell$ mutually disjoint maximal commuting operator
classes form a set of $\ell$ mutually unbiased bases.\\
\el

\subsection{Unextendibility of MUBs and operator classes}

A set of MUBs $\{ B_1, B_2, \ldots ,$ $B_{\ell} \}$ is called {\em unextendible} if there does not exist another basis that is unbiased with respect to the bases $B_1,\ldots,B_{\ell}$.

The correspondence between MUBs and maximal commuting operator classes gives rise
to a weaker notion of unextendibility, based on unextendible sets of such classes.

\bd[Unextendible sets of operator classes \cite{Woot}] {\rm 
A set of mutually disjoint maximal commuting classes $\{ \mC_1, \mC_2, \ldots , \mC_{\ell} \}$ of operators drawn from a unitary basis $\mU$ is said to
be {\em unextendible} if no other maximal class can be formed out of the remaining operators in
$\mU \setminus (\{\bI\} \cup \bigcup_{i = 1}^{\ell} \mC_i)$.}
\ed

The eigenbases of such an unextendible set of classes form a weakly unextendible set of MUBs,
as defined below.

\bd[Weakly unextendible sets of MUBs \cite{Woot}] {\rm
Given a set of MUBs $\{B_1, B_2,$ $\ldots, B_{\ell} \}$ that are realized as common eigenbases of a set of $\ell$ operator classes comprising operators
from $\mU$, the set $\{ B_1, B_2,\ldots, B_{\ell}\}$ is {\em weakly unextendible} if there does not exist another unbiased
basis that can be realized as the common eigenbasis of a maximal commuting class of operators
in $\mU$.}
\ed

\bigskip
\section{Symplectic polar spaces and the Pauli group}

Consider the projective space $\PG(2N - 1,d)$ of dimension $2N -
1$, $N \geq 2$, over the field $\mathbb{F}_d$ with $d$ elements, $d$ an odd prime. Let $F$ be a
{non-degenerate} symplectic form of $\PG(2N - 1,d)$. 
For $F$ one can choose the following canonical bilinear form \cite{Hirsch}:

\[ (X_0Y_1 - X_1Y_0) + (X_2Y_3 - X_3Y_2) + \cdots + (X_{2N - 2}Y_{2N - 1} - X_{2N - 1}Y_{2N - 2}).          \]

Then the {\em symplectic polar space} $\mW_{2N -
1}(d)$ consists of the points of $\PG(2N - 1,d)$ together with all
totally isotropic spaces of $F$ \cite{Hirsch}. Here, a {\em totally isotropic subspace} is 
a linear subspace of $\PG(2N - 1,d)$ that vanishes under $F$.

One can also define
this space in the underlying $2N$-dimensional vector space
$V(2N,d)$ over $\mathbb{F}_d$ using a {non-degenerate} 
alternating bilinear form (which induces a symplectic form  on the
projective space).\\

\br[Number of points]{\rm
Note that $\vert$ points of $\mW_{2N - 1}(d)\vert = \frac{\vert V(2N,d)\vert - 1}{d - 1} = d^{2N - 1} + d^{2N - 2} + \cdots + 1$.\\
}\er

In the following proposition, $[.,.]$ denotes the commutator relation in the group $\hP$.

\begin{proposition}[K. Thas \cite{Appl}]
\label{prop}
\begin{itemize}
\item[{\rm (i)}]
The derived group $\hP' = [\hP,\hP]$ equals the center $Z(\hP)$ of $\hP$.
\item[{\rm (ii)}]
We have $Z(\hP) = \langle \omega\bI\rangle$, so that $\vert Z(\hP)\vert = d$.
\item[{\rm (iii)}]
$\hP$ is nonabelian of exponent $d$.
\item[{\rm (iv)}]
We have the following short exact sequence of groups:
\[    1 \mapsto Z(\hP) \mapsto \hP \mapsto V(2N,d) \mapsto 1.            \]
\end{itemize}
\end{proposition}

\begin{remark}
{\rm Note that if $d = 2$, $\hP$ is nonabelian of exponent $4$.
}
\end{remark}

\bigskip
Now denote the natural map $\hP \mapsto V(2N,d)$ by an overbar. Then the commutator

\[ [.,.]: V(2N,d)\times V(2N,d)  \mapsto \langle \omega\mathbb{I}_{d^N}\rangle:  (\overline{v_1},\overline{v_2}) \mapsto [ \overline{v_1},\overline{v_2}] = [v_1,v_2] \]
\noindent
defines a {non-degenerate} alternating bilinear form on $V(2N,d)$, so defines a symplectic polar space $\mW_{2N - 1}(d)$.
{Here the derived group $\hP'$ is identified with the additive group of $\mathbb{F}_d$}.\\

\begin{theorem}[\cite{Appl}]
\label{conn1}
Two elements of $\mathbb{P}^{\times}$ commute if and only if the corresponding points of $\mW_{2N-1}(d)$ are collinear. 
In other words, the commuting structure of $\hP$ (and $\mathbb{P}$) is governed by  that of the symplectic polar space $\mW_{2N - 1}(d)$.
\end{theorem}

Applying this result, one can easily construct sets 
of MUBs of maximal size $\ell + 1$ using the symplectic geometry \cite{Appl}. \\

\bigskip
\section{Unextendable mutually unbiased bases and Pauli classes}

In this section, we explain in detail the correspondence between Pauli classes and generators of symplectic polar spaces of \cite{Appl}. It has the same proof as Theorem \ref{conn1}, but we make the relation between (unextendible) commuting Pauli classes and the generators more explicit.

\begin{theorem}[General connecting theorem]
\label{ps-umub}
Two elements of $\mathbb{P}^{\times}$ commute if and only if the corresponding points of $\mW_{2N-1}(d)$ are collinear. 
In other words, the commuting structure of $\hP$ (and $\mathbb{P}$) is governed by  that of the symplectic polar space $\mW_{2N - 1}(d)$.
As a corollary, ``complete'' partial spreads of $\mW_{2N - 1}(d)$, correspond to unextendible sets of operator classes in the Pauli group.
\end{theorem}

We indicate the proof in several steps.

{ Let $d$ be any prime and $N \in \mathbb{N} \setminus \{0,1\}$.
Let $\mS$ be a {\em partial spread} of  $\mW_{2N - 1}(d)$, i.e., a set of $(N-1)$-dimensional isotropic subspaces which are two by two disjoint. Let $M + 1$ be the number of elements in $\mS$, and note that $M + 1 \leq d^N + 1$ (equality holds when $\mS$ is a spread).
{Then $\mS$ corresponds to a set of mutually unbiased bases in the associated $d^{N}$-dimensional Hilbert space, in the following way.} 
 
 \begin{itemize}
 \item[Step 1]
 To $\mS$ corresponds a set of $M  + 1$ subgroups $H_i$, $i\in
 \{0,1,\ldots,M\}$, 
 of $\hP$ of size $d^{N + 1}$ which mutually (two by two) intersect (precisely) in $Z(\hP)$.
 \item[Step 2]
In each $H_j$ one chooses $d^N - 1$ elements $H_j^k$ ($k = 1,2,\ldots,d^N - 1$) which are not contained in $Z(\hP)$, so that no two such elements are in the same $Z(\hP)$-coset.
\item[Step 3]
Then $\mU(\mS) := \{H_\alpha^\beta \vert \alpha \in \{0,1,\ldots,M\}$, $\beta \in \{1,2,\ldots,d^N - 1\}\}$ is a set of commuting unitary classes. 
\item[Step 4] 
If $\mS$ is a {\em complete} partial spread of $\mW_{2N - 1}(d)$, that is, if $\mS$ is not strictly contained in {\em another} partial spread, then 
 $\mU(\mS)$ is unextendible, and the corresponding set of MUBs is weakly unextendible of size $M + 1$.
 \end{itemize}
 
 In particular, this construction applies when $\mU(\mS)$ is a set of Pauli operators (each $Z(\hP)$-coset contains precisely one Pauli operator).
 
 \subsection{The bijection $\rho$}
 
 Let $\mG(\mW_{2N-1}(d))$ be the set of generators of $\mW_{2N-1}(d)$, and let $\mC(\mathbb{C}^{d^N})$ be the set of commuting classes of 
 Pauli operators (of size $d^N - 1$).
 Note that from the above, it follows that we have a {\em bijection} 
 \begin{equation}
 \rho: \mC(\mathbb{C}^{d^N}) \longrightarrow \mG(\mW_{2N-1}(d))
 \end{equation}
 which sends an element $\mU \in \mC(\mathbb{C}^{d^N})$ to a generator, following the scheme explained above. It is indeed a
 bijection: to each generator corresponds a unique maximal abelian subgroup $A \leq \hP \leq \mathbf{U}_{d^N}(\mathbb{C})$ as above (and conversely), and each $Z(\hP)$-coset in this subgroup contains precisely one  Pauli operator. Together, the set of (nontrivial) Pauli operators in $A$ form one commuting class of Pauli operators of size $d^N - 1$, that is,  one element of $\mC(\mathbb{C}^{d^N})$.

 \subsection{Prime dimension}
 
 {Now let $N = 1$ (i.e., prime dimension) and $d \ne 2$. Then Proposition} \ref{prop} {tells us that $\hP$ is a group of size $d^3$ and exponent $d$, and its center has size $d$ | in other words,
 $\hP$ is {\em extra-special}. }
{ In $\hP$ one can now choose subgroups $H_j^i$ as above, and again \cite{Band} applies. If $d = 2$, the result is well known (but it can also be derived as above).}

\bigskip
\section{The case $d$ prime, $N = 2$ | small and large examples}

If $d = p$ is a prime and $N = 2$, the corresponding symplectic polar space is $\mW_3(p) =: \mX$, with ambient projective space $\PG(3,p)$, 
and it has two types of linear subspaces which are completely contained in $\mX$, namely points of $\PG(3,p)$, and (projective) lines.

\subsection{Grids and point-line duals}

Before proceeding, we explore some synthetic properties of $\mW_3(p)$ which will makes things easy below.

Let $\mP$ be the point set of $\mW := \mW_3(p)$, $\mL$ its line set, and let $\bI$ be the (symmetric) ``incidence relation'' on $(\mP \times \mB) \cup
(\mB \cup \mP)$ which says that $x \bI L$ ($L \bI x$) if and only if the point $x$ is on the line $L$. Then this point-line geometry is a {\em generalized quadrangle} \cite{FGQ}, and a very deep and extensive theory exists on these structures. Note that each line contains $p + 1$ points and 
that on each point there are $p + 1$ lines. Also, recall the following  defining projection property for generalized quadrangles: if $x$ is a point and 
$X$ a line not containing $x$, there is a unique line $Y \I x$ which meets $X$ (and then in a unique point).

Now consider the point-line geometry $\mQ$ with line set $\mB$, point set $\mP$ and the same incidence relation $\bI$ | the so-called {\em point-line dual}  of $\mW_3(p)$. Then by \cite[3.2.1]{FGQ}, $\mQ$ is isomorphic to the point-line geometry of an orthogonal quadric $\mQ(4,p)$ in $\PG(4,p)$. 
Moreover, if $p = 2$, $\mQ$ and $\mW$ are isomorphic \cite[3.2.1]{FGQ}. 

Let ``$\perp$'' denote the orthogonality relation in $\mW_3(p)$, and let $V, W$ be arbitrary lines which do not meet. Then $\{ V, W\}^{\perp}$
consists of $p + 1$ lines which are mutually disjoint. If $p$ is odd, ${(\{ V, W\}^{\perp})}^{\perp}$, the set of lines which meet all lines of 
$\{ V, W \}^{\perp}$, is $\{ V, W\}$. If $p = 2$, there is a third line $X$ besides $V, W$ in this set. The set of points on the lines of $\mR_1 := \{V, W\}^{\perp}$,
which is the same as the set of points on the lines of $\mR_2 := {(\{ V, W\}^{\perp})}^{\perp}$, forms a $(3 \times 3)$-grid $\mG$, and the aforementioned line sets $\mR_1, \mR_2$ are the  reguli of this grid.

Also, an easy counting exercise shows that 
all lines of $\mW_3(2)$ have at least one point in common with the point set of $ V \cup W \cup X$. (All these properties can essentially be found in 
\cite[Chapter 3]{FGQ}.)

\begin{center}
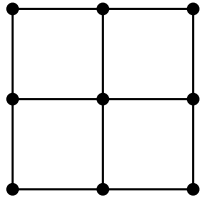
\begin{figure}
  \begin{tikzpicture}[style=thick, scale=1.2]
\foreach \x in {1,2,3}{
\fill (\x,1) circle (2pt);}
\foreach \x in {1,2,3}{
\fill (\x,2) circle (2pt);}

\foreach \x in {1,2,3}{
\fill (\x,3) circle (2pt);}



\draw (1,1) -- (2,1);
\draw (2,1) -- (3,1);

\draw (1,2) -- (2,2);
\draw (2,2) -- (3,2);

\draw (1,2) -- (1,1);
\draw (2,2) -- (2,1);
\draw (3,2) -- (3,1);

\draw (1,3) -- (2,3);
\draw (2,3) -- (3,3);

\draw (1,2) -- (1,3);
\draw (2,2) -- (2,3);
\draw (3,2) -- (3,3);
\end{tikzpicture}

\caption{A $(3 \times 3)$-grid.}
\end{figure}
\end{center}

\subsection{Antiregularity}

If $\ell$ is any odd prime power, we will use the fact that $\mW_3(\ell)$ has no $(3 \times 3)$-grids. This is a corollary
of a property called ``antiregularity,'' and can be found in \cite[3.3.1(i), dual]{FGQ}.

\medskip
\subsection{The case $p = 2$}

We start with giving an alternative and very short proof of Theorem 1 of \cite{Woot} using the connection between Pauli classes and 
partial spreads of symplectic polar spaces.

\bt[Mandayam et al. \cite{Woot}]
Given three Pauli classes $\mC_1,\mC_2,\mC_3$ belonging to a complete set $\mS$ of classes in $d = 4$, there exists exactly one more maximal
commuting class $\mC$ of Pauli operators in $\mC_1 \cup \mC_2 \cup \mC_3$.  The class $\mC$ together with the remaining two classes $\mC_4$ and $\mC_5$ of $\mS$ forms a unextendible set of Pauli classes, whose common eigenbases form a weakly UMUB of order $3$.
\et

{\em Proof}.\quad
Interpret $\mS$ in $\mW_3(2)$; then by Theorem \ref{ps-umub} to the $\mC_i$ correspond lines $L_i$ of $\mW_3(2)$ ($i = 1,\ldots,5$), and they form a spread. 
Consider the lines $L_1,L_2,L_3$. Then either there is precisely one line $L$ of $\mW_3(2)$ meeting them all, or there are three such lines.
In the latter case the lines $L_1,L_2,L_3$ form a regulus of a $(3 \times 3)$-grid, and as we have seen any line of $\mW_3(2)$ meets 
the point set of such a grid, leading to the fact that $L_1,L_2,L_3$ would not be extendible to a spread, contradiction. So we are in the fomer case, 
and the class $\mC$ corresponding to $L$ is the one of the statement. 
The set $\mC,\mC_4,\mC_5$ obviously is unextendible, since extending it with a class $\widetilde{\mC}$ would mean than the corresponding line 
$\widetilde{L}$ be contained in the point set of $L_1 \cup L_2 \cup L_3$, implying that there would be another line besides $L$ meeting all of $L_1,L_2,L_3$, contradiction.
\eop \\

We now give a short proof of another result of \cite{Woot}, namely Theorem 5 of that paper.

\bt[Mandayam et al. \cite{Woot}]
Given an unextendible set of three Pauli classes $\mC_1,\mC_2,\mC_3$  in $d = 4$, the nine operators in $\mC_1 \cup \mC_2 \cup \mC_3$
can be partitioned into a different set of three maximal commuting classes $\mC_1',\mC_2',\mC_3'$ such that each $\mC_i'$ has precisely one operator in common with each $\mC_j$, $i, j \in \{ 1,2,3\}$.
\et

{\em Proof}.\quad
Let $L_1, L_2, L_3$ the lines corresponding to $\mC_1, \mC_2, \mC_3$ in $\mW_3(2)$; we have seen that either one or three lines are 
contained in $\{L_1,L_2,L_3\}^{\perp}$; in the latter case, an easy counting argument shows that all lines of $\mW_3(2)$ intersect with
$L_1 \cup L_2 \cup L_3$, so suppose we are in the former case, and let $\{ L\} := \{L_1,L_2,L_3\}^{\perp}$. Then each point on $L$ is incident
with precisely one line besides $L$ and not in $\{L_1,L_2,L_3\}$. By the projection property of generalized quadrangles, there are six lines
different from $L_1, L_2, l_3$ which meet  the six points of $L_1 \cup L_2 \cup L_3$ not on $L$ in precisely two points. So the total
number of lines meeting $L_1 \cup L_2 \cup L_3$ is $13$, and $\{L_1,L_2,L_3\}$ is indeed extendible (since there are $15$ lines in total).
\eop \\

In the next subsection, we will see a general approach for constructing unextendible Pauli classes in $\mathbb{C}^{d^2}$ with $d$ a prime number, starting from a complete set.
As a corollary, we will obtain yet another proof for the result of Mandayam et al.

\subsection{General case}

\bt[Existence of unextendible Pauli class sets for $d$ prime, A]
\label{exprime}
For each prime $d = p$ there exists  an unextendible set of Pauli classes $\mS$ of size $d^2 - d + 1$ or $d^2 - d + 2$ in $\mathbb{C}^{d^2}$.
The common eigenbases form a weakly UMUB of order $d^2 - d + 1$ or $d^2 - d + 2$.
\et
{\em Proof}.\quad
As before, we pass to $\mW_3(p)$.
Let $\mT$ be any spread of $\mW_3(p)$. 

Now let $U$ be any line of $\mW_3(p)$ which is not contained in $\mT$; then there are precisely $p + 1$ lines in $\mT$ which 
hit $U$ (each in exactly one point), due to the fact that the lines of $\mT$ partition the point set of $\mW_3(p)$. Call this line set 
$\mT_U$. Now consider the line set 
\begin{equation}
\mT(U) := \mT \setminus \mT_U \cup \{U\}.
\end{equation}
Note that $\vert \mT(U) \vert = p^2 - p + 1$.
If it is not a complete partial spread, there is at least one other line $R$ of $\mW_3(p)$ not meeting any line of $\mT(U)$, 
and as a point set it clearly must be contained in the point set ``of'' $\mT_U$. And then all lines of $\mT_U$ meet both 
$U$ and $R$. If yet another line $R'$ would exist that extends $\mT_U \cup \{R\}$, $R'$ would also be met by the lines of $\mT_U$ |
in other words, $R' \in \mT_U^{\perp}$ while $\mT_U =  U^{\perp} \cap R^{\perp}$. As we have seen, this is not possible, so 
only at most one line $R$ can be added. 

Translating back to Pauli classes gives the desired result.
\eop \\

It is easy to see that both cases of Theorem \ref{exprime} can occur.

As we have not used the fact that $p$ is prime, we can translate immediately to symplectic polar spaces over any finite field.

\bc
For each prime power $\ell$ there exists  a complete partial spread  $\mS$ of $\mW_3(\ell)$ of size $\ell^2 - \ell + 1$ or $\ell^2 - \ell + 2$. \eop 
\ec

\medskip
\br
{\em
For $\ell$ even, we have seen this result in the literature (see, e.g. \cite{cimrak} and the references therein) | it would be safe to attribute this result to folklore though. We presume the odd case is somewhere as well, but the way of proving 
is needed below, so it is included anyhow for the sake of completeness.
}
\er

One could apply the technique in the proof of Theorem \ref{exprime} multiple times to obtain examples with less elements. And indeed, this 
works quite well, as we will demonstrate now. We will work immediately in $\mW_3(\ell)$, and will not restrict ourself only to the prime case. So $\ell$ is a prime power. We {\em do} ask that $\ell$ is odd | it will be used in the proof.\\

Let $\mS$ be a {\em classical spread} of $\mW_3(\ell)$ | by this, we mean a spread which in the point-line dual $\mQ(4,\ell)$ corresponds to 
an elliptic quadric. Take any two lines $L, M$ in $\mS$, and consider the set $\mX = \{ X_0,X_1,\ldots,X_\ell \} := \{L,M\}^{\perp}$; it consists of $\ell + 1$ mutually  disjoint lines which are not in $\mS$. Now for each $X_i \in \mX$, define $\mS_i$ to be the set of $\ell + 1$ lines of $\mS$
meeting $X_i$. As explained in Appendix B of this paper, for each $\mS_i$ there is precisely one more line $\widetilde{X_i} \ne X_i$
which meets each line of $\mS_i$. Clearly, this line must be in $\mX$, so we can denote $\widetilde{X_i}$ by $X_{\widetilde{i}}$.

Now the following properties are immediate:

\begin{itemize}
\item[(a)]
$\widetilde{(\cdot)}: \{0,1,\ldots,\ell\} \longrightarrow \{ 0,1,\ldots,\ell\}$ is an involution, so that $\vert \{ \mS_0,\mS_1,\ldots,\mS_{\ell}\} \vert = (\ell + 1)/2$; 
\item[(b)]
for $\mS_i  \ne \mS_j$, we have that $\mS_i \cap \mS_j = \{L,M\}$.\\
\end{itemize} 

For the sake of convenience, we re-write the set $\{\mS_0,\mS_1,\ldots,\mS_{\ell}\}$ as $\{\mS_0,\mS_1,\ldots,\mS_{(\ell - 1)/2}\}$. For each
$j \in \{0,1,\ldots,(\ell - 1)/2\}$ we have that $\{ X_j,X_{\widetilde{j}}\}^{\perp} = \mS_j$.

\medskip
\bt
Let $\ell$ be an odd prime power. Then for any $k = 0,1,\ldots,(\ell - 3)/2$, there exist complete partial spreads of size 
$\ell^2 - (k + 1)\ell + (3k + 1)$ in $\mW_3(\ell)$.
\et
{\em Proof}.\quad
Let $k$ be as in the statement, and consider any subset $R$ of $\{0,1,\ldots,(\ell - 1)/2\}$ of size $k + 1$; for simplicity, we consider w.l.o.g.
the set $\{0,1,\ldots,k\}$. Then define the following set:
\begin{equation}
\mS_R := \mS \setminus (\cup_{u \in R}\mS_u) \cup_{v \in R}(\{X_v,X_{\widetilde{v}}\}).
\end{equation}

It is straightforward to see that $\mS_R$ is a partial spread of size $\ell^2 - (k + 1)\ell + (3k + 1)$. As for completeness, suppose 
we could enlarge $\mS_R$ with some line $U$ to another partial spread. As $\mS$ is a spread, $U$ must be contained 
in $\cup_{u \in R}\mS_u$, and it cannot be contained in $\mS$ nor $\mS_R$. By the Pigeon Hole Principle, some 
$\mS_w$ must have at least three lines meeting $U$ since $U$ has $d + 1$ points and $k + 1 < \frac{d + 1}{2}$ (in case $U \in \{ L,M\}^{\perp}$, 
one does not need the Pigeon Hole Principle). However, this implies the existence of a $(3 \times 3)$-grid, 
contradiction. \eop \\

For each odd prime power $\ell$ the bounds appear to be new (up to some small coincidences). For fixed $\ell$, we obtain complete partial spreads of respective sizes
\begin{equation}
\ell^2 - \ell  + 1, \ell^2 - 2\ell + 4, \ell^2 - 3\ell + 7, \ldots, \frac{\ell^2}{2} + 2\ell - \frac{7}{2}.  
\end{equation}

Translating back to Pauli operators, we obtain the next result.\\

\bt[Existence of unextendible Pauli class sets for $d$ prime, B]
\label{exprime}
For each odd prime $d = p$ and any $k = 0,1,\ldots,(d - 3)/2$ there exists  an unextendible set of Pauli classes $\mS$ of size $d^2 - (k + 1)d + (3k + 1)$ in $\mathbb{C}^{d^2}$.
The common eigenbases form a weakly UMUB of order $d^2 - (k + 1)d + (3k + 1)$. \eop \\
\et

The construction has many variations, all using roughly the same ideas, and all giving similar (but not the same) bounds.
We will come back to these variations in a forthcoming paper.\\

\bigskip
\section{Solution of Conjecture \ref{Wootconj}}

Motivated by Theorem \ref{Wootmain1}, the following conjecture is then stated in \cite{Woot}.

\bcc[Mandayam et al. \cite{Woot}]
\label{conj2}
Given $\ell/2 + 1$ maximal commuting Pauli classes $\mC_1,\mC_2,\ldots,\mC_{\ell/2 + 1}$ belonging to a complete set $\mS$ of classes in $\ell = 2^N =: d^N$, $N \in \mathbb{N} \setminus \{0,1\}$, there exists exactly one more maximal
commuting class $\mC$ of Pauli operators in $\cup_{1 \leq i \leq \ell/2 + 1}\mC_i$.  The class $\mC$ together with the remaining classes of $\mS$  forms an unextendible set of Pauli classes of size $\ell/2 + 1$, whose common eigenbases form a weakly UMUB of order $\ell/2 + 1$.
\ecc

In this section we will show that this conjecture is true {\em if and only if $N = 2$ or $N = 3$}. In fact, we will consider the conjecture in any characteristic $d$ (i.e., for any prime $d$), and show that it is true if and only if $d = 2$ and $N = 2$ or $N = 3$.

Translated to the geometric setting, we obtain: ``given $2^{N-1} + 1$ elements $\alpha_1,\alpha_2,\ldots,\alpha_{2^{N - 1}+1}$ belonging to a spread $\mS$ of the polar space $\mW_{2N - 1}(2)$,
there exists exactly one more generator $\chi$ which is completely contained in the union of these elements, such that $\chi$ together with 
the remaining elements of $\mS$ constitutes a complete partial spread.''

Note that the situation implies that $\chi$ meets each $\alpha_j$, $j = 1,2,\ldots,2^{N-1}+1$.

We will replace $d = 2$ by any prime $d$, and consider the same situation (immediately in the geometric setting). We will also slightly generalize the statement by replacing ``exactly one'' by ``at least one.''

So let $\mS$ be a spread of $\mW_{2N - 1}(d)$, $d$ a prime. We assume that the conjecture above is true (in the more general setting).

First suppose  that $\mU$ and $\mU'$ are different subsets of $\mS$, both of size $d^{N - 1} + 1$. Let $\alpha$ be a generator
which meets all elements of $\mU$ and is covered by these elements, and let $\alpha'$ be a generator
which meets all elements of $\mU'$ and is covered by them. Then $\alpha \ne \alpha'$.

In the next counting argument, we will use the fact that the number of generators of $\mW_{2N - 1}(d)$ is $(d^N + 1)(d^{N - 1} + 1)\cdots(d + 1)$.
Per subset of $\mS$ of size $d^{N - 1} + 1$, by the conjecture we have at least one generator meeting all of its elements, and covered by them. Such a generator 
is never contained in $\mS$. So we have that 
\begin{equation}
\label{eq1}
C_{\vert \mS \vert}^{d^{N - 1} + 1}\cdot 1 + \vert \mS \vert \leq (d^N + 1)(d^{N - 1} + 1)\cdots(d + 1).
\end{equation}
Here,
\begin{equation}
C_{\vert \mS \vert}^{d^{N - 1} + 1} := \frac{(d^N + 1)!}{(d^{N -1} + 1)!(d^N - d^{N - 1})!}.
\end{equation}
(Note that equality should hold in (\ref{eq1}) in the ``precisely one statement.'')

Now (\ref{eq1}) is equivalent to
\begin{equation}
\frac{(d^N + 1)d^N\cdots(d^N - d^{N - 1} + 1) + (d^{N - 1} + 1)!(d^N + 1)}{(d^{N - 1} + 1)! (d^N + 1)(d^{N - 1} + 1)\cdots(d + 1)} \leq 1,
\end{equation}
or, slightly simplyfied:

\begin{equation}
\frac{((d^N + 1)/(d^{N - 1} + 1))(d^N/d^{N - 1})\cdots ((d^N - d^{N - 1} + 1)/(1)) + (d^N + 1)}{(d^N + 1)(d^{N - 1} + 1)\cdots(d + 1)} \leq 1.
\end{equation}

Now note that 
\begin{equation}
((d^N + 1)/(d^{N - 1} + 1))(d^N/d^{N - 1})\cdots ((d^N - d^{N - 1} + 1)/(1)) \geq \frac{d^N + 1}{d^{N - 1} + 1}\cdot d^{d^{N - 1}},
\end{equation}
and that 
\begin{equation}
(d^N + 1)(d^{N - 1} + 1)\cdots(d + 1) \leq d^{N + 1}d^N\cdots d^2 = d^{N(N + 3)/2}.
\end{equation}

Observe that if for some value $N = M$, we have 
\begin{equation}
d^{d^{M - 1}} \geq d^{M(M + 3)/2}, 
\end{equation}
then the same inequality holds  for all $M' \geq M$. 

This is already enough to conclude with a contradiction for $d \geq 5$; $d = 3$ and $N \geq 3$; and $d = 2$ and $N \geq 6$.
The cases $(d,N) = (3,2), (2,5), (2,4)$ all yield a contradiction when substituted in (\ref{eq1}); the substitution $(d,N) = (2,2)$, 
which is precisely the case of $\mW_3(2)$ which was already studied before, leads to equality in (\ref{eq1}), as does the 
substitution $(d,N) = (2,3)$,  which is the case of $\mW_5(2)$. \eop \\

Note that the cases $(d,N) = (2,2), (2,3)$ are precisely those handled in \S 3, Theorem 1 and \S 3, Theorem 3 of \cite{Woot}.\\

In the next section we will formulate and discuss variations on Conjecture \ref{Wootconj}; to that end, we first try to generalize Theorem \ref{exprime}.\\

\bigskip
\section{Existence of maximal Pauli classes}

Before proceeding, let us first introduce a simple lemma about complete ``partial spreads'' of general incidence structures. 
Let $\Gamma = (\mE,t,T)$ be a triple, with $T = \{ 0,1,\ldots,n \}$, $n \in \mathbb{N}^\times$, and $t$ a surjective function from the set $\mE \ne \emptyset$
to $T$. For each $i \in \{ 0,1,\ldots,n\}$, put $\mE_i := t^{-1}(i)$, and call its elements the elements {\em of type $i$}. 
So
\begin{equation}
\mE = \cup_{i \in T}\mE_i,\ \ \mbox{and}\ \ \vert \mE \vert \geq \vert T \vert.
\end{equation}
In particular, we call elements of $\mE_0$ ``points.'' We now assume that for $i > 0$, every element of $\mE_i$ is a subset of $\mE_0$. 
This is a natural assumption: we see each ``$i$-space'' (= element of type $i$) as a point set.\\

 An {\em $i$-spread} of $\Gamma$ is a partition of $\mE_0$ in elements of type $i$. {\em Complete $i$-spreads} are introduced naturally as above.
 
 \bp
 \label{propconst}
 Let $\mS$ be an $i$-spread of $\Gamma$. Let $\chi$ be an element of type $i$ which is not contained in $\mS$, and let $\mS_{\chi}$ be the 
 subset of elements of $\mS$ which meet $\chi$ in at least one point. Note that $\mS_{\chi}$ induces a partition of the points of $\chi$. 
 Then if we cannot find a set $\mT$ of elements of type $i$ such that
 \begin{itemize}
 \item[{\rm {\bf C.1}}]
 each element of $\mT$ is a subset of the point set
 \begin{equation}
 \Omega(\mS,\chi): = \cup_{U \in \mS_{\chi}}U; 
 \end{equation}
 \item[{\rm {\bf C.2}}]
 the elements of $\mT$ partition $\Omega(\mS,\chi)$,
 \end{itemize}
 we have that $\mS \setminus \mS_{\chi} \cup \{ \chi\}$ cannot be completed to an $i$-spread of $\Gamma$.
 \ep
 {\em Proof}.\quad
 If $\mS \setminus \mS_{\chi} \cup \{ \chi\}$ {\em could} be completed to an $i$-spread $\mS'$ of $\Gamma$, 
 $\mS'$ must have elements which all are subsets  of $\Omega(\mS,\chi)$, and which partition $\Gamma$.
 \eop \\

If $\ell$ is the maximum number of elements of type $i$ contained in $\Omega$ as subsets and which are two by two disjoint, 
the number of elements in a maximal partial $i$-spread containing $\mS \setminus \mS_{\chi} \cup \{ \chi\}$ is at most 
$\vert \mS \vert - \vert \mS_{\chi} \vert + \ell$. (Note that $\ell \geq 1$ as $\chi$ itself is in $\Omega(\mS,\chi)$.)\\

\br[Back to the prime case]
{\rm
Note that the first part of Theorem \ref{exprime} is an application of the construction method of Proposition \ref{propconst} (with $\chi = L$).
}
\er

\medskip
\subsection{$\mU$-Sets}

Motivated by Proposition \ref{propconst}, a {\em $\mU$-set} with {\em carrier} $\chi$ is a set $\mS_{\chi}$ of mutually disjoint generators of $\mW_{2N - 1}(d)$ which all meet some generator $\chi \not\in \mS_{\chi}$ such that 
\begin{equation}
\chi \subset \cup_{Y \in \mS_{\chi}}Y,
\end{equation}
and such that $\cup_{Y \in \mS_{\chi}}Y$ cannot be partitioned by a partial spread $\mP$ of generators which includes $\chi$.

Note that the number of elements of an $\mU$-sets is not uniquely determined by $N$ and $d$. (One $\mU$-set could also have different carriers.)

\bp[Existence of UMUBs, I]
\label{exUMUB}
The existence of $\mU$-sets implies the existence of complete partial spreads which are not spreads, that is, of unextendible sets of Pauli classes.
\ep
{\em Proof}.\quad
Let $\mS_{\chi}$ be a $\mU$-set. If $\mS_{\chi}$ is not contained in a spread, then we are done, so suppose it {\em is} contained in some 
spread $\mS$. Then by Proposition \ref{propconst} we have that $\mS \setminus \mS_{\chi} \cup \{ \chi\}$ cannot be completed to a spread. 
\eop \\

Note that this proposition can also be applied to general incidence geometries.\\

For the rest of this section, we suppose $d$ is an odd prime.

Before proceeding, recall that a spread $\mS$ (of generators) of $\mW_{2N - 1}(d)$ is {\em regular} if the following property is satisfied: if for every three distinct elements $\alpha, \beta, \gamma$ in $\mS$,  $\mL$ is the set of lines of $\PG(2N - 1,d)$ which meet each of $\alpha, \beta, \gamma$, 
then there are $d - 2$ further elements of $\mS$ which meet every line in $\mL$. It is well known that every symplectic polar space
has regular spreads.

Now let $\mR$ be a regular spread of $\mW_{2N - 1}(d)$. Take a generator $\chi$ which meets some $\alpha \in \mR$ in a space
of dimension $N - 2$ (and note that this is possible), and let $\mR_{\chi}$ be the set of elements in $\mR$ which meet $\chi$. Note that $\vert \mR_\chi \vert = d^{N - 2} + 1$. Now consider a generator $\beta \ne \chi, \alpha$ which contains $\chi \cap \alpha$, and which is disjoint from the elements in $\mR_{\chi} \setminus \{\alpha\}$. (For the existence of such a generator, see Appendix A.)
Then because $\mR$ is a regular spread, one notes that $\mS_{\chi} := \mR_{\chi} \setminus \{\alpha\} \cup \{\beta\}$ is a $\mU$-set.
For, suppose  that $\cup_{Y \in \mS_{\chi}}Y$ can be partitioned by a partial spread $\mP$ of generators which includes $\chi$. Let $\gamma \in \mP \setminus \{\chi\}$ contain some point $b$ of $\beta$; then $\gamma \cap \beta = \{b\}$. Let $B$ be any line in $\gamma$ containing $b$;
then $B$ meets $d + 1$ different elements of $\mS_{\chi}$, one of which is $\beta$. As $d \geq 3$, the fact that $\mR$ is a regular spread
implies that $b \in \alpha$, contradiction.\\

In the next theorem, we prove that unextendible sets of Pauli classes of $\mathbb{C}^{d^N}$ always exist (ignoring possible sizes completely), that is, that complete partial spreads which are not spreads always exist in $\mW_{2N - 1}(d)$. This fact is not necessarily true
for general incidence geometries which have $i$-spreads (using the nomenclature of above): consider for instance an incidence geometry for which the elements of type $i$ precisely form {\em one} $i$-spread. So although the existence of complete partial spreads is probably seen as folklore, we see the need to formally write it down.

\bt[Existence of UMUBs, II]
Every Hilbert space $\mathbb{C}^{d^N}$, with $d$ an odd prime and $N$ a positive integer, contains weakly unextendible sets of Pauli classes.
\et
{\em Proof}.\quad
Translated to $\mW_{2N - 1}(d)$, we need to show that the latter geometry always contains complete partial spreads which are not spreads.
So take a regular spread $\mR$. Consider a generator $\chi$ as above, and construct the $\mU$-set $\mS_{\chi} := \mR_{\chi} \setminus \{\alpha\} \cup \{\beta\}$. Now apply Proposition \ref{exUMUB}.
\eop \\

\br{\rm
Note that if $\mR_{\chi}$ in the proof of the previous theorem is such that there does not exist a generator besides $\chi$ which is contained in 
$\cup_{\alpha \in \mR_{\chi}}\alpha$, then 
\begin{equation}
\vert \mR \setminus \mR_{\chi} \cup \{ \chi\} \vert = d^{N} - d^{N - 1} + 1.
\end{equation}

In the special case $d = 2$, we would end up with an unextendible set of Pauli classes of size $2^{N - 1} + 1$.}\\
\er

\subsection{Reformulation of  Conjecture \ref{conj2}}

We have seen that Conjecture \ref{conj2} is only true when $N = 2$ or $N = 3$. On the other hand, there seems to be some evidence that 
the bound of that conjecture could be attained (see, e.g., the previous remark). 
So we reformulate the conjecture as follows | we will do it in geometric terms, over all fields $\mathbb{F}_{\ell}$ with $\ell$ a prime power, but
again, for the applications in Quantum Information Theory, one takes $\ell$ to be prime. 

\bc
For each prime power $\ell$ and positive integer $N \geq 2$, there exists a spread $\mS$ of $\mW_{2N - 1}(\ell)$ and 
a generator $\chi$ not in $\mS$, such that  $\mS \setminus \mS_{\chi} \cup \{ \chi\}$ is a complete partial spread of size
 $d^{N} - d^{N - 1} + 1$.
\ec 

When $d = 2$, one obtains the same bound as in Conjecture \ref{conj2}. 

We hope to come back to this conjecture in the near future.\\

\bigskip
\section{``Galois MUBs''}

When $d = 2, 3, 5, 7$ or $11$, there exist extremely exotic examples of unextendible sets of Pauli classes of size $d^2 - 1$ in $\mathbb{C}^{d^2}$.
(Details, constructions and references can be found in \cite{SDW-KT}.) We propose to call the corresponding sets of MUBs ``Galois MUBs,''
because they are all related to exotic $2$-transitive representations of special linear groups, as was first noted by Galois (see also \cite{SDW-KT}).
They are extremely special amongst Pauli classes of $\mathbb{C}^{d^2}$, $d$ a prime, or even {\em all} Hilbert spaces, due to the following result.

\bt[See \cite{FGQ}, \S 2.7]
Let $\Gamma$ be a generalized quadrangle of finite thick order $(s,s)$, and let $\mC$ be a complete partial spread of $\Gamma$.
If $\Gamma$ is not contained in a spread of $\Gamma$, then 
\begin{equation}
\vert \mC \vert \leq s^2 - 1.
\end{equation}
\et

As we have seen that the points and lines 
of any $\mW_3(d)$ form a generalized quadrangle, this result applies to $\mW_3(d)$ and hence also to Pauli classes in $\mathbb{C}^{d^2}$.

\bc
A set of commuting Pauli classes of size $d^2$ in $\mathbb{C}^{d^2}$, $d$ a prime, is never unextendible.
\eop \\
\ec

\br{\rm
The aforementioned examples in $d = 2, 3, 5, 7, 11$ are the only known examples which effectively reach the $(d^2 - 1)$-bound, and conjecturally they are the only ones.  Geometrically, they also satisfy very extreme properties, which rightly translate to Pauli classes.
Much more details on the geometric structure of partial spreads of size $s^2 - 1$ in generalized quadrangles of order $(s,s)$ can be found
in the author's paper \cite{KTarcs}. 
}
\er

The next theorem, taken from the author's paper \cite{KTarcs}, says that when $d = 2$, up to isomorphism there is only one complete partial spread of size $3 = 2^2 - 1$ in $\mW_3(2)$.

\bt[\cite{KTarcs}]
Up to isomorphism there is only one complete partial spread of size $3$ in $\mW_3(2)$.\\
\et

\bc
Up to isomorphism, there is only one unextendible set of Pauli classes of size $3$ in $\mathbb{C}^4$. 
\eop \\
\ec

\br[On isomorphisms]
{\rm 
Of course, one needs to specify {\em what} isomorphisms between unextendible sets of Pauli classes {\em are}. Because of the General Connecting Theorem
(and the bijection $\rho$), we propose to say that unextendible sets of Pauli classes $\mU$ and $\mU'$ in $\mathbb{C}^{d^N}$ are {\em isomorphic} if 
there exists an automorphism of $\mW_{2N - 1}(d)$ which maps the complete partial spread $\mS(\mU)$ corresponding to $\mU$, to   the complete partial spread $\mS(\mU')$ corresponding to $\mU'$. This is a natural notion of ``isomorphism,'' since automorphisms of $\mW_{2N - 1}(d)$
preserve collinearity of points, so also the commuting of operators at the level of Pauli operators. (One could also define isomorphisms through
the general Pauli group. On the other hand, such automorphisms induce automorphisms of $\mW_{2N - 1}(d)$ anyhow, while the converse is {\em not} true. So one misses (many) maps which should be considered as isomorphisms in this way.)
}
\er

\newpage
\section{Conclusion}

The geometry underlying the space of the generalized Pauli operators/matrices characterizing $N$ $d$-level quantum systems, with $N \geq 2$ and $d$ any prime, is that of the symplectic polar space of rank $N$ and order $d$, $\mW_{2N-1}(d)$. 

Using this connection, we have derived a short proof of a recent result of \cite{Woot} on the unextendibility of MUB sets in $\mathbb{C}^4$ (their Theorem 1). Moreover, we generalized this result for all $d =$ square of a prime, and  presented a construction of a class of maximal partial spreads of $\mW_3(\ell)$ for any odd prime power $\ell$, attaining new bounds in generically every case,
which rightly translates to UMUBs in the case $\ell$ is a prime.
We also gave a very short proof of Theorem 5 of \cite{Woot}.

We then considered Conjecture 1 of \cite{Woot} which conjecturally generalizes the aforementioned result of \cite{Woot} to {\em any} dimension and showed that it is true {\em if and only if $N = 2$ or $N = 3$}.

We then indicated that an alternative version of the conjecture might be true, and described several new construction techniques to obtain  weakly unextendible sets of MUBs. 

Finally, we discussed a special kind of weakly unextendible sets of MUBs, called ``Galois MUBs,'' which attain an optimal bound in relation to being unextendible.

\newpage
\appendix
\section{Properties of (symplectic) polar spaces}

Consider the space $\mW_{2N - 1}(d)$, $d$ a prime, $N \geq 2$. (We restrict ourselve to the prime case because that's the class which translates to Pauli operators, but everything works when $d$ is a prime power as well.)\\

\subsection{}
Let $\gamma$ be a generator, and $x$ a point not in $\gamma$. Then a well-know property (of general polar spaces) | see e.g. \cite[p.137, (c)]{FIG} | says that there is a unique generator on $x$ which meets $\gamma$ in an $(N - 2)$-space, $\gamma(x)$. (If ``$\perp$'' is the orthogonality relation coming from the associated alternating form, then $\gamma(x) = x^{\perp} \cap \gamma$.)
Now let $\gamma$ and $\gamma'$ be disjoint  generators. Then it is not hard to see that for every $(N - 2)$-space $\delta$ in $\gamma'$, there is precisely one point $y \in \gamma$
such that $\langle y,\delta \rangle$ is a generator ($y = \gamma \cap \delta^{\perp}$).
So the map
\begin{equation}
\mu: \gamma \longrightarrow \gamma': x \longrightarrow \gamma'(x)
\end{equation}
is a bijection between the points of $\gamma$ and the hyperplanes of $\gamma'$. \\

\subsection{}
Now let $\alpha$ be an $(N - 2)$-space contained in $\mW_{2N - 1}(d)$; it is well-known that there are $d + 1$ generators $g_0,\ldots,g_{d}$ containing $\alpha$.
Let $\beta$ be a generator disjoint from $\alpha$. By the surjectivity above, it follows that some $g_i$ has to intersect $\beta$, and then necessarily in one point.\\

\subsection{Structure of spreads}
Let $\mS$ be a spread of $\mW_{2N - 1}(d)$, $d$ a prime, $N \geq 2$. 
Let $\alpha = g_0 \in \mS$, and let $\tau$ be  an $(N - 2)$-space in $\alpha$.  Let $g_1,\ldots,g_d$ be the other generators containing $\tau$.
By the previous paragraph, each element of $\mS \setminus \{\alpha\}$ meets some $g_i$ ($i \ne 0$) in precisely one point. 
And conversely, each point of $g_j \setminus \tau$ is contained in precisely one spread element. Indeed,
\begin{equation}
(\vert \mS \vert - 1)\cdot 1 = d^N = d\cdot d^{N - 1} = \sum_{i = 1}^d\#(\mbox{points}\ \ \mbox{of}\ \ g_i \setminus \tau).
\end{equation}

\newpage
\section{Some more properties of $\mW_3(d)$}

As in the first appendix, for the applications in Quantum Information Theory considered here, one wants to think of $d$ as being 
prime, but everything holds when $d$ is a prime power as well.  What we {\em do} ask is that $d$ is odd.\\

Let $\mS$ be a classical spread of $\mW_3(d)$; point-line dualize to obtain $\mQ(4,d)$ | $\mS$ becomes an elliptic quadric, denoted $\mO$.
Now let $x$ be a point of $\mQ(4,d)$ not contained in $\mO$. As usual, let ``$\perp$'' denote the orthogonality relation 
associated to the defining quadratic form of $\mQ(4,d)$ (say, corresponding to the variety with equation $X_0^2 + X_1X_2 + X_3X_4 = 0$).
Then $x^{\perp} \cap \mO$ is a conic section, and since $d$ is odd, there is precisely one other point $y \not\in \mO$ for which
\begin{equation}
y^{\perp} \cap \mO = x^{\perp} \cap \mO.
\end{equation}
Note that the latter expression is equal to $\{x,y\}^{\perp}$.

Going back to $\mW_3(d)$ (i.e., dualizing again), we obtain that if $X$ is a line of $\mW_3(d)$ not in $\mS$, and 
$\mS_X$ is the set of $d + 1$ lines in $\mS$ which meet $X$, then there is precisely one other line $Y$ not in $\mS$ such that
\begin{equation}
\mS_X = \{X,Y\}^{\perp} = X^{\perp} \cap Y^{\perp} = \mS_Y.
\end{equation}

\end{document}